\documentclass[aps,amsfonts,twocolumn,nofootinbib]{revtex4-1}
\usepackage{epsfig}
\usepackage{graphicx}
\usepackage{slashed}
\usepackage{amsmath}
\usepackage{amsbsy}
\usepackage{bm}
\usepackage{color}
\usepackage{epsfig}
\usepackage{ulem}
\newcommand{\ee}{\end{equation}}
\newcommand{\bb}{\begin{equation}}
\newcommand{\eqb}{\begin{eqnarray}}
\newcommand{\eqf}{\end{eqnarray}}

\def\p{\mathbf{p}}

\newcommand{\1}{{\'{\i}}}

\def\1{\'{\i}}

\def\1{\'{\i}}
\begin{document}
\title{Cosmic Neutrino Background as a Ferromagnet }
\author{Paola Arias}
\email{paola.arias@fis.puc.cl}
\address{Departamento de F\'isica, Pontificia Universidad Cat\'olica de Chile, Casilla 306, Santiago 22, Chile}

\author{Jorge Gamboa}
\email{jgamboa55@gmail.com}
\address{Departmento de F\1sica, Universidad de Santiago de Chile, Casilla 307, Santiago, Chile}

\author{Justo L\'opez-Sarri\'on}
\email{justinux75@gmail.com}
\address{Departmento of F\1sica, Universidad de Santiago de Chile, Casilla 307, Santiago, Chile}

\begin{abstract}
If cosmic background neutrinos interact very weakly with each other, through spin-spin interactions, then   they may have experienced a phase transition, leading to a ferromagnetic ordering. The small magnetic field resulting from ferromagnetic ordering  -- if present before galaxy formation -- could act as a  primordial seed of the magnetic fields observed in several galaxies. Our findings suggest that the magnetization could occur in the right epoch, if the exchange boson of neutrino-neutrino interaction is a massless boson beyond the Standard Model, with a coupling constant of $2.2\times 10^{-13} \left(\frac{m_\nu}{10^{-4}\,\rm{eV}}\right)^2<g<2.3\times 10^{-7}$. The estimation of the magnetic seed is  $2.3 \times 10^{-27}\,\rm{G}\lesssim B_{\rm CNB}\lesssim 6.8\times 10^{-10}G$.
\end{abstract}
\pacs{PACS numbers:}
\date{\today}
\maketitle
\section{Introduction}
There are two untested hypothesis, that if confirmed would allow us to better understand the first moments of the early universe: they are the detection of the theorized Cosmic Neutrino Background (CNB) \cite{CNB} and the generation mechanism of a primordial magnetic seed \cite{Grasso:2000wj, rubins,review1, review2}.

The detection of the CNB is  hard because, besides  the well known weak interactions of neutrinos, the CNB is decoupled from the matter content of the universe and has a temperature of about  $T_{0\nu}\sim 1.9$K.  Therefore,  observing the CNB seems like an impossible quest. 

In order to look for indirect consequences of the CNB, it is needed to identify the key processes involved. One possibility is that neutrino-neutrino interactions \cite{bialynicki} could lead to an observable phenomenon, but it is expectable that -- in order for them to be effective enough to leave an observable trace -- the exchange particle of the process should not be the usual $W$ or $Z$ boson, but a particle beyond the Standard Model \cite{Bardin:1970wq}. {These} interactions {can indeed} play an important role, {at some epoch after the neutrino decoupling.} 

If the latter is true, we find spin-spin interaction as the dominant one in the CNB, provided that the momentum transferred to the new boson is small.  This last fact means that the processes  $\nu\nu \rightarrow \nu\nu$ -- at this low energy limit -- induce a spin effective interaction $\bf{S} \cdot \bf{S}$, such as the spin models in statistical mechanics \cite{Wipf:2013vp}. In other words, the CNB undergoes a phase transition which is responsible of the magnetization of the background.

 On the other  hand, magnetic fields up to $\sim \mu$G have been observed in galaxies. {{Their unknown origin has prompt several ideas about generation mechanisms }}\cite{zeldovich}. Some scenarios assume that a primordial magnetic seed gets adiabatically compressed when protogalactic cloud collapse \cite{Grasso:2000wj, rubins,review1, review2}. Other proposals point to an astrophysical mechanism for the generation of these fields, a Biermann battery effect \cite{Kulsrud:1992rk,naoz}. It is unknown whether one of these mechanisms is actually  responsible of the observed magnetic fields. Independently of the mechanism,  the existence of primordial fields has an impact in the development of cosmology, such as the success of Big Bang Nucleosynthesis, structure formation, etc. \cite{Grasso:2000wj,rubins,review1,review2}.

In this paper we propose that neutrinos from the CNB interact with each other via the exchange of an intermediate $X$ boson, much lighter than $Z_0$.  
{{These neutrinos acquire a}} spontaneous magnetization from their spin-spin interaction and if this magnetization was  present before the galaxy formation,   would provide a mechanism for a primordial magnetic seed.

The article is organized as follows: in section II we present an effective model for the interaction $\nu \nu \rightarrow \nu \nu$ and we write {{the corresponding}} effective Hamiltonian, using a Breit approximation for small  transferred momentum between neutrinos and the new  boson. We argue why spin-spin interaction should be the dominant one among the neutrinos in the CNB, and we write the critical temperature of the magnetic phase transition. Further, we impose constraints to this new interaction, {{firstly for a low mass scale}} and secondly for a {{large mass scale}}.  In section III we estimate the magnitude of the magnetic field generated in the ferromagnetic phase transition and section IV contains the conclusions. 
\section{{Effective interaction}}
Let us start by considering the coupling between {Dirac} neutrinos to a generic light neutral  boson ($X_\mu$)
\bb
\mathcal L=- {g} \bar \nu \gamma_\mu \nu X^\mu + \frac{1}{2} M^2_X X^2, 
\ee
where a sum over neutrino species is assumed.  {{Formally, we will refer to $M_X$ as the mass of the exchange boson, and $g$ the effective coupling constant of a secret sector with neutrinos. In this sense, the vector field $X_\mu$ parametrizes the secret neutrino interactions, but does not necessarily represents a fundamental particle.}}

Defining the neutral current as $J_\mu= \bar \nu \gamma_\mu \nu$, the effective Hamiltonian density can be written as a four-neutrino interaction, namely
\bb
\mathcal H= \frac{g^2}{M_X^2}  J_\mu J^\mu.
\ee
The next step  is to study the process $\nu \nu \to \nu \nu $  by using the Breit approximation and assuming an almost vanishing transferred momentum  to the new vector boson, namely $\bf{q}=\p_{\rm ini}-\p_{\rm fin}$. Thus,  the two-body Breit potential \cite{piitaevski} is  
\bb
 H_I= - \frac{g^2}{M_X^2}  n_\nu \sum_{\{i,j\}} \,{\bf S}^{(i)} \cdot {\bf S}^{(j)}, \label{heisen}
\ee
where indices  $i, j$  are  particles fixed indices and the bracket notation $\{i,j\}$ stands for nearest neighbors particle interaction. The parameter $n_\nu$ is the neutrino number density. 

In (\ref{heisen}) we have not written the Coulomb, dipolar, quadrupolar and spin-orbit interactions because they are negligibly small compared with spin-spin interaction. 

The Hamiltonian (\ref{heisen}) is obtained by computing at  tree-level the scattering amplitude in $1/c$ powers, {{although, it is not difficult to show that this term is the leading effect even in a relativistic case}}. In this quantum field theory context we get  
\bb 
\frac{g^2}{M_X^2} \sum_{i} \delta (x_i -x_{i+1})\,{\bf S}^{(i)} \cdot {\bf S}^{(i+1)} \rightarrow \frac{g^2}{M_X^2} \, n_\nu  \sum_{i} {\bf S}^{(i)} \cdot {\bf S}^{(i+1)}, \nonumber 
\ee 
and therefore the delta-function corresponds to  the neutrino density.  
Note that the minus sign in the RHS  (\ref{heisen}) appears naturally, implying  the interaction is ferromagnetic.

We stress spin-spin interaction is the dominant one, because of the transferred momentum to the new vector boson is small. This turns the interaction exclusively as a contact one.

The case of interest to us is when there is a phase transition (at some critical temperature $T_c$)  and then CNB acquires a spontaneous magnetization. 

In order to investigate this, we use the mean field approximation, {\it i.e} the interaction Hamiltonian is replaced by 
\bb
 H_I= -J \langle S\rangle \sum_i\, S_i ,
\ee
where $J=\frac{g^2}{M_X^2} \, n_\nu$, $\langle S\rangle$ is the mean value of the spin,  and the condition for a phase transition is
\bb
\frac{J}{T}\gtrsim1.
\ee
By identifying $J=T_c$, the critical temperature, the phase transition condition reads $T_c>T$.

Replacing the value of $J$ and using that the neutrino density of the CNB scales with the redshift as $n_\nu= n_0 \left(1+z\right)^3$, where $n_0$ is the present neutrino density, $n_0\sim 56$ cm$^{-3}$,  and the temperature scales as $T=T_{0\nu} \left(1+z\right)$, with $T_{0\nu}\sim 1.9$K, we find
\bb
{\left(1+z\right)^2} \gtrsim \frac{T_{0\nu}{M_X^2}}{{g^2}  n_{0}}.
\label{zup}
\ee

The above equation can be used if an upper bound for the redshift is applied. 

On cosmological grounds, it is natural to require that the new interaction gets enhanced at an epoch, $z_i$, after the CNB decoupling, or if present before that, should not be the predominant one.  This reasoning leads us to set an upper limit of $z_i<z_{\nu \rm {dec}}$, where the decoupling of neutrinos from the thermal bath is known to be around $z_{\nu \rm{dec}}\sim 10^9$ \cite{Cyr-Racine:2013jua}.

On the other hand, has been found in ref.\cite{Basboll:2008fx}, that a re-coupling of the CNB -- if existing -- should not be present at recombination era ($z_{\rm rec}\sim 1500$), where neutrinos are free-streaming \cite{DeBernardis:2008ys}.  We will assume that at recombination epoch, neutrinos free-stream, namely, neutrino-neutrino interaction rate, $\Gamma_{\nu\nu\rightarrow \nu\nu}$,  does not exceed the cosmic expansion rate, $H_{\rm rec}$ at this redshift
\bb
\Gamma_{\nu\nu\rightarrow \nu\nu}(z_{\rm rec}) < H_{\rm rec}.
\label{hubble}
\ee
 From these two requirements we can find if a new vector boson, interacting only with neutrinos,  could lead to a ferromagnetic ordering of the latter, which is not in conflict with observations.
\subsection{Light vector boson exchange ($M_X\lesssim m_\nu$) }
Let us first consider the case of a very light boson, such that $M_X$ is of the order of neutrino mass, {\it{ i.e.}} $M_X \sim m_\nu$. In this case eq.~(\ref{zup}) becomes
\bb
(1+z)^2> \frac{m_\nu^2 T_{0\nu}}{n_{0} g^2},
\ee
By imposing that the interaction neutrino-hidden boson turns on at an epoch after neutrino decoupling, $z_i< 10^9$, we get
\bb
2.2\times 10^{-13}\left(\frac{m_\nu}{10^{-4}\,\rm{eV}}\right)^2 \leq g.
\ee
On the other hand, {{to respect the free-streaming of neutrinos}} at the epoch of recombination, we require that at $z_i=1500$, the interaction should be negligible.

At recombination era,  the universe was matter dominated, so the Hubble parameter is given by $H_{\rm rec}=100\, \mbox{km}^{-1}\, \mbox{Mpc}^{-1}\left(\Omega_Mh^2\right)^{1/2}\left(z_{\rm rec}+1\right)^{3/2}$. Where $\Omega_Mh^2$ is the cosmic matter density and is given by  $\Omega_Mh^2=0.134$. The Hubble parameter at this redshift takes the value $H_{\rm rec} =4.5\times 10^{-29}$ eV. 

On the other hand,  neutrino-neutrino interaction rate is given in  this (nearly massless) case by $\Gamma_{\nu\nu\rightarrow \nu\nu}=n_\nu \langle \sigma v\rangle$, where  $\langle \sigma v\rangle \sim g^4n_\nu/\langle s\rangle $, and the free-streaming requirement, eq.~(\ref{hubble}), reads
\bb
g< 2.3 \times 10^{-7}.
\ee
Combining both limits we find that spontaneous magnetization of the CNB is possible, {at a redshift in between, $1500< z_i<10^{9}$}, if the exchange  boson of neutrino-neutrino interactions is  a nearly massless vectorial boson, with a constant coupling of
\bb
2.2\times 10^{-13} \left(\frac{m_\nu}{10^{-4}\,\rm{eV}}\right)^2\leq g \leq2.3\times 10^{-7}.
\label{limit}
\ee 
\subsection{Massive vector boson exchange ($M_X \gg m_\nu$)}
Secondly, we will consider a massive vector boson, such $M_X\gg m_{\nu}$. In the massive case, the constrained parameter is the Fermi-like constant $G_X\equiv \frac{g^2}{M_X^2}$. From the requirement that the interaction turns on at a temperature, $T_c$, after neutrino decoupling we get, using eq.~(\ref{zup})
\bb
G_X \geq 4.8 \times 10^{13}\, G_F,
\ee
where, $G_F\sim 10^{-23}$eV$^{-2}$,  is the Fermi constant.

Meanwhile, the requirement that the interaction should be weak enough to let neutrinos free-streaming at recombination era, $\Gamma_{\nu\nu\rightarrow \nu\nu}/H_{\rm rec}<1$, gives
\bb
G_X\leq 6.4\times 10^{10}\, G_F,
\ee
where we have used that the interaction rate is $\Gamma_{\nu\nu\rightarrow \nu\nu}= G_X^2 T^2 n_\nu$. 

Putting both region of interest together, we find that spontaneous magnetization of the CNB mediated by a massive  boson is not possible at the epoch before recombination, because the two required constraints do not complement each other
\bb
4.8 \times 10^{13}\, G_F \leq G_X \leq 6.4\times 10^{10}\, G_F.
\ee
Also, a vector boson, {in any of these two limits}, has been ruled out from several astrophysical tests \cite{Bilenky:1999dn}.
\section{Estimating the  Magnetic Seed Field}
A rough estimation of the magnetic field generated at the phase transition can be made by approximating each neutrino as a spin. An upper limit in the magnetic field would be to assume that all species of neutrinos undergo the ferromagnetic phase transition. With this  assumption, the  magnetic seed field is
\bb
B_{\rm CNB}= \mu_n n_\nu.
\ee
Where $\mu_\nu$ is the neutrino magnetic moment, predicted to be $\mu_\nu\sim 10^{-19}\mu_B$ \cite{momento,Fujikawa:1980yx}. Replacing in the above equation, and using our redshift window of $1500<z_i<10^{9}$, we get an estimate  of
\bb
2.3 \times 10^{-27}\,\rm{G}\lesssim B_{\rm CNB}\lesssim 6.8\times 10^{-10}G.
\ee
The maximal correlation length for the magnetic field would be the Hubble radius at the time of generation, $H^{-1}(z_i)$. Therefore, the maximal correlation length of the field generated by the spontaneous magnetization ranges from $\lambda_{\rm dec} \sim 3.4$npc at neutrino decoupling era, to $\lambda_{\rm rec} \sim 100$kpc at recombination.

The cosmological magnetic field  damps with the scale factor as $a^{-2}$,  so a quick check tell us that an observed field of $B\sim 1\mu$G in galaxy clusters, corresponds to a primordial magnetic field of $B_{\rm prim}\sim 1$nG at the epoch $z\sim 1100$ \cite{Yamazaki:2012pg}, which seems in agreement with our result.

\section{Discussion and summary}
In this paper we have proposed a mechanism by which the Cosmic Neutrino Background could become observable in a new physics framework. Such a mechanism is produced by an enhancement of neutrino-neutrino interactions after their decoupling from the thermal bath, in the early universe. 

We have found that if the scattering has a small momentum transfer, $|{\bf{q}}|$, the dominant  interaction in the CNB is a ferromagnetic spin-spin coupling. Such interaction --if existing-- {{should have a coupling strength given by ($\ref{limit}$),  and a low mass mediator.}}

As a result of the coupling, a ferromagnetic phase in the CNB appears at a temperature before structure formation ($z\sim 10 $), and thus, the resulting magnetization can act as a primordial magnetic seed, and later be amplified by a dynamo mechanism. 

Our results are not in contradiction with limits found in several related proposals~\cite{Kolb:1987qy, Bardin:1970wq, Manohar:1987ec,Dicus:1988jh,Bilenky:1992xn,Bilenky:1994ma,Masso:1994ww}. Although the idea proposed here is the result of observing CNB neutrinos as a spin system, this interpretation is also consistent and complementary with the secret interactions studied in  the above mentioned papers. It is remarkable that the behavior of neutrinos at very low scale of energy can be very unconventional and a detailed study, very likely, should provide unexpected phenomena .

Finally we have made an estimation of the magnitude of the magnetic field generated by the CNB, assuming all neutrinos from the three species undergo the ferromagnetic phase transition. The strength of the magnetic field depends on the epoch at which the transition took place. For a redshift between $1500< z_i< 10^9$, (corresponding {{to a temperature in the range of}} MeV-eV), the magnetic field could be in the range of $2.3 \times 10^{-27}\,\rm{G}\lesssim B_{\rm CNB}\lesssim 6.8\times 10^{-10}G.$
\\

\section*{Acknowledgements}

We would like to thank Professors C. A, Garcia-Canal, F. A. Schaposnik, A. Ringwald, A. Reisenegger and A. Wipf for valuable comments. This work was supported by grants from FONDECYT-Chile 11121403 {{and anillo ACT 1102}}  (P.A.),  1130020 (J.G.) and  1100777, 1140243 (J.L-S.) and DICYT 041131LS.

\end{document}